# Using unsupervised machine learning to quantify physical activity from accelerometry in a diverse and rapidly changing population

Christopher B Thornton[1*], Niina Kolehmainen[1,2], and Kianoush Nazarpour[3]

[1] Population Health Sciences Institute, Faculty of Medical Sciences, Newcastle University, UK

[2] Great North Children's Hospital, Newcastle upon Tyne NHS Hospitals Trust, UK

[3] Institute for Adaptive and Neural Computation, School of Informatics, The University of Edinburgh, UK

* Corresponding Author:

E-mail: chris.thornton@newcastle.ac.uk

Abstract

Accelerometers are widely used to measure physical activity behaviour, including in children. The traditional method for processing acceleration data uses cut points to define physical activity intensity, relying on calibration studies that relate the magnitude of acceleration to energy expenditure. However, these relationships do not generalise across diverse populations and hence they must be parametrised for each subpopulation (e.g., age groups) which is costly and makes studies across diverse populations and over time difficult. A data-driven approach that allows physical activity intensity states to emerge from the data, without relying on parameters derived from external populations, and offers a new perspective on this problem and potentially improved results. We applied an unsupervised machine learning approach, namely a hidden semi-Markov model, to segment and cluster the accelerometer data recorded from 279 children (9-38 months old) with a diverse range of physical and social-cognitive abilities (measured using the Paediatric Evaluation of Disability Inventory). We benchmarked this analysis with the cut points approach calculated using the best available thresholds for the population. Time spent active as measured by this unsupervised approach correlated more strongly with measures of the child's mobility, social-cognitive capacity, independence, daily activity, and age than that measured using the cut points approach. Unsupervised machine learning offers the potential to provide a more sensitive, appropriate, and cost-effective approach to quantifying physical activity behaviour in diverse populations, compared to the current cut points approach. This, in turn, supports research that is more inclusive of diverse or rapidly changing populations.

# Introduction

Participation in physical activity is widely considered to be beneficial for all people. This includes young children for whom physical activity is known to promote development and positive health outcomes (1), while spending time sedentary has been shown to result in poor sleep (2). Accurately measuring participation in physical activity by members of this age group, including those with non-typical developmental trajectories, is an important step in the development of interventions seeking to facilitate participation.

Accelerometers are increasingly used to measure the wearer's physical activity. They record the acceleration of the body part to which they are attached, providing an objective record of how much movement has occurred. This raw acceleration recording can then be processed to extract the time spent by the wearer in a range of physical activity intensity categories, such as "sedentary" (SED), "light physical activity" (LPA), and "moderate to vigorous physical activity" (MVPA). The traditional method used to process the raw acceleration trace into these categories, known as the cut points method, applies a threshold to the volume of acceleration recorded in each epoch (3). This threshold is calibrated in a detailed lab-based study, where the wearer's energy expenditure is measured at the same time as the acceleration, so that the cut point thresholds indicate the volume of acceleration at which the participant would be expected to expend a predefined level of energy. The energy levels of interest are usually calculated as a ratio of the energy expended while at rest – known as a Metabolic Equivalent (METs) – and for children are typically <1.5 METs for SED, 1.5-3 METs for LPA, and >3 METs for MVPA (4). As the relationship between acceleration volume and energy expenditure depends upon the physical abilities, body size, and movement patterns of the child (5) calibration must be performed for each subpopulation. This results in different cut points derived as children age and develop(6) and for children with different movement capabilities (7). Using the cut-points approach is therefore challenging, if not impossible, when the population under study is diverse in their physical capabilities or when they are rapidly changing, such as in a longitudinal study of young children.

Recently, machine learning approaches have been increasingly applied to the analysis of accelerometer data. Some have sought to train supervised machine learning techniques to recognise activity types (8,9) while others have used unsupervised techniques to categorise activity based on the intensity and direction of movement (10,11). This latter approach promises to segment and cluster the acceleration data according to its intensity, without parameters derived from external populations, and to offer a potentially more appropriate method for determining engagement in physical activity in longitudinal population studies of young children.

In the present study we adopt a hidden semi-Markov model (HSMM) (12) to segment and cluster accelerometer data in 279 toddlers. We wish to show that this method is a more appropriate approach in a rapidly developing and highly heterogeneous population. For comparison, we process the accelerometer data according to the traditional cut-points approach, using the best available parameters for this population. To evaluate which approach provides the most clinically relevant measure, we compare both approaches to how each child scores on the Paediatric Evaluation of Disability Inventory Computer Adaptive Test (PEDI-CAT), an established assessment of developmental capacities with strong psychometric properties across age groups and populations.

## Materials and Methods

### Ethics statement

The data presented was collected as part of the ActiveCHILD project (NIHR ICA-SCL-2015-01-003)(13,14) The study had NHS Research Ethics Committee and Health Research Authority (UK) approvals (NHS IRAS 218313, 17-NE-0051), and the design drew on the Nuffield Ethics guidance for health research involving children(15).

Data collection

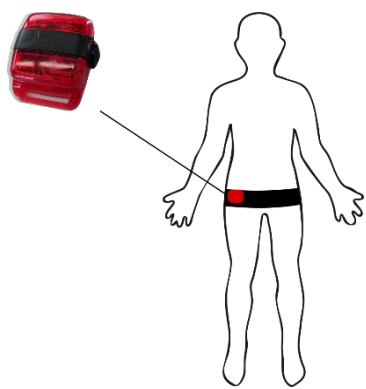

*Figure 1: Shows the accelerometer (the Actigraph wGT3X-BT) and how it was instructed to be worn.*

Young children (n=279, aged between 9 and 36 months) were recruited through the universal healthy child pathway (i.e. health visitors) and specialist children's health services (e.g. neonatology, community paediatrics, paediatric physiotherapy) in thirteen areas in England, UK. Children recruited through specialist services (42%, n=118/278) were purposefully oversampled to ensure coverage of children with a range of development trajectories. Table 1 shows the key characteristics of the sample; further details are available elsewhere(14).

*Table 1: Shows the distribution of ages and recruitment pathways of the participants.*

|  | Total number of children | 279 |
|---|---|---|
| Recruitment (n=278) | Health Visitor | 58% (160) |
|  | Other paediatric specialist | 42% (118) |
| Sex (% female) (n=218) |  | 56% (123) |
| Age (months) (n=277) | 5 - 10 | 5% (13) |
|  | 10 – 15 | 27% (76) |
|  | 15 - 20 | 18% (50) |
|  | 20 -25 | 17% (47) |
|  | 25 - 30 | 22% (61) |
|  | 30 - 35 | 10% (28) |
|  | 35 - 40 | 1% (2) |
| Mobility as described by the clinician (n=202) | Walks without support | 80% (162) |
|  | Uses walking aid | 10% (21) |
|  | Moves with support only (e.g. parent carrying, buggy) | 9% (19) |
| Cognitive development (Clinician's Assessment) (n=203) | Unable to comment on the child's cognitive capacity | 19% (39) |
|  | No concerns about the child's cognitive development | 68% (139) |

| | There are concerns about this child's cognitive development or learning (e.g. the child is below educational level, the child receives support for learning) | 9% (18) |
|---|---|---|
| | The child has a global developmental delay established as part of a multidisciplinary or medical assessment | 3% (7) |

To collect data, families were provided with pre-prepared accelerometer packs. The pack contained a pre-programmed accelerometer threaded on a flexible, waist-worn belt, and instructions(16) for the parent. The device was set to record two days after the parent received the pack, and parents were encouraged to let their child play with the accelerometer on these pre-recording days in order to familiarise with the device. On the night before the first recording day, parents were asked to place the accelerometer beside the child's bed, to prompt them to put it on first thing in the morning. The parents were then asked to encourage the child to wear the accelerometer for seven days, except while bathing or showering, swimming, or in bed. At the end of the recoding period, parents were asked to post the accelerometer to the research team. In return of the accelerometer, parents were sent a feedback sheet of their child's activity and the child was sent a small toy as a reward. The ActiGraph GT3X+ was used, which has been previously found to be acceptable and feasible to use in under5s (17) with physical limitations as well as typically developing children. The device was worn around the waist, set to record at 100 Hz, and set to capture record all movement that lasted at least one second.

Motor and social-cognitive capacity were measured using the Pediatric Evaluation of Disability Inventory - Computer Adaptive Test (PEDI-CAT), completed by a parent responding for their child. The PEDI-CAT software utilises Item Response Theory (IRT) statistical models to estimate a child's abilities from a minimal number of the most relevant items or from a set number of items within each domain. It focuses on the child's ability to perform activities in a manner that is effective given their abilities(18) and does not require the child to perform the activity in a standardised manner.

Accelerometer data pre-processing

The accelerometer data was processed using the Python programming language, with the gt3x module used to read the raw accelerometer signal from file. From this we extracted the 10 second mean of the vector magnitude of body acceleration, calculated as the Euclidian Norm Minus One (ENMO). This is calculated as shown in the equation below, where $accx, accy, accz$ are the acceleration along the three orthogonal axes. We have used the ENMO because it automatically corrects for the contribution of gravity by subtracting one gravitational unit from the overall magnitude and has been shown to perform well in segmenting accelerometery according to activity intensity (19).

$$ENMO = \{max\sqrt{(acc_x^2 + acc_y^2 + acc_z^2)} - 1, 0\}.$$

As the devices recorded movement at all times, regardless of whether the device was being worn or not, the first step in our data processing was to detect when the device was worn by the child in the morning and when it was last taken off at night. To do this we segmented the recording into days starting and ending at 4am, calculated the activity to be the mean of the acceleration in 10 second epochs, then took the first worn time to be the first epoch of non-transient activity, and the last worn time to be the last epoch of non-transient activity. We defined non-transient activity as activity recorded for at least 150 seconds out of every 1000 seconds. These parameters were determined empirically by applying a range of parameters to a sample of recordings and selecting those that corresponded best with human selected wear times. We then detected any non-wear time during the day as any continuous period of zero activity longer than one hour. As many of the children had worn the device for more than seven days, or had worn the device more sporadically, we then selected the seven continuous days that provided the maximum amount of wear time. A recording day was only included in the analysis if it contained at least five hours of wear time; and the overall recording could only be included if it contained at least three days of suitable recordings(20).

Training the HSMM

The HSMM model allowed us to segment and cluster our accelerometry traces according to the magnitude of acceleration. It can segment the trace into periods of similar acceleration, then cluster each segment into one of a number of hidden states. Hidden states are defined by their parameters (in this case parameters describing their acceleration distribution and duration distribution). The HSMM extends the better-known Hidden Markov Model (HMM) by explicitly modelling the duration of time spent in a state. This is modelled as a Poisson distribution and the observations (acceleration magnitudes) modelled by a Gaussian distribution. The trained model also contains a transition probability matrix, defining the likelihood of transitioning from one state to another. As the parameters for these distributions are learned in a Bayesian manner, the HSMM allows the hidden states to emerge solely from the data under study, unlike the traditional cut points method whose parameters must be derived from studies of external (and possibly unrepresentative) populations. We used the pyhsmm Python package (21) to train the HSMM, which implements a Hierarchical Dirichlet Process Hidden semi-Markov Model . The HSMM model requires us to specify a number of hyperparameters that influence the outcomes of the learning process as well as the computational resources required. We set the maximum state duration to 360 ten second epochs or 60 minutes. We set the maximum number of states to be six – this would allow a long duration and short duration state for each of the cut point categories (SED, LPA, MVPA). A condition for early stopping was specified to be when the Hamming distance between two consecutive iterations was less than 0.05, and a maximum of 20 iterations was used. The random seed was set to zero.

Applying the cut-points approach

Many of the cut points specified in the literature are in units of accelerometry counts rather than raw accelerations. To use the most appropriate cut points for our population, we processed the raw accelerometry into accelerometry counts, with an epoch of one second, using the ActiLife software provided by ActiGraph. We then labelled each one second epoch using the following rules:

- Sedentary activity (SED) as less than eight counts per second (based on the estimate of 40 counts per 5 seconds (7)).

- Light physical activity (LPA) as more than eight counts per second (7) and less than 28 counts per second - based on the an estimate of 420 counts per 15 seconds (22,23).

- Moderate to vigorous physical activity (MVPA) as more than 28 counts per second (22,23).

We then found bouts of continuous activity using a custom function written in Python based on the getBout.R function of the GGIR package (24). This allowed 10% of an LPA bout or SED bout and 20% of an MVPA bout to be an interruption.

# Results

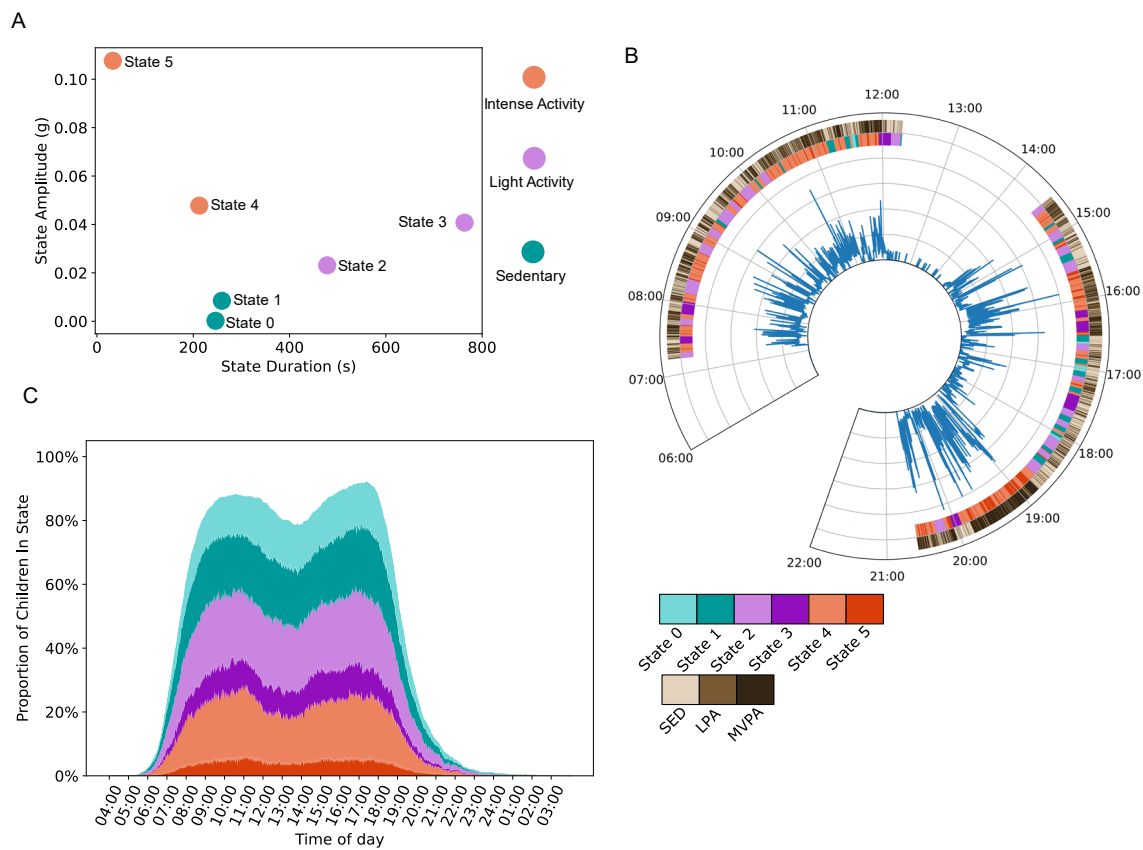

*Figure 2: (A) Shows the parameters of each state in the HSMM. The colour corresponds to the group they have been placed in. (B) Shows a sample accelerometry trace from one day of recording (central trace) for one child and the corresponding classification according to the HSMM (inner circle) and traditional cut points approach (outer circle). The legend below indicates the states or cut point categories represented by each colour. (C) shows the proportion of children in each HSMM state throughout the day. The legend to the right indicates the colours corresponding to each state.*

The cut points approach and HSMM approach were applied to data from 279 children, with each child contributing between three and seven days of accelerometry recordings. Figure 2 (A) shows the parameters of the six hidden states of the HSMM after training. The state duration is the λ of the Poisson distribution that models the duration of each activity state, the state amplitude is the mean of the Gaussian distribution that models the acceleration magnitude of each state. Here we can see that states with a high acceleration mean also have a shorter duration. This reflects the underlying feature that high intensity physical activity cannot be maintained for the same length of time as low intensity activity. We also see that the states can be clustered around three groups, high intensity-short duration (states 4 and 5), low intensity-long duration (states 2 and 3), and very low intensity (states 0 and 1) – the colour of each state reflects the group we have assigned it to. To facilitate comparison with the traditional physical activity intensity categories (SED, LPA, MVPA) we compare SED with states 0 and 1, LPA with states 2 and 3, and MVPA with states 4 and 5.

Figure 2 (B) shows an example accelerometry trace for one day of activity along with an annotation indicating the HSMM states assigned to each segment as well as the classical cut points categories (SED, LPA, MVPA). Here we can see an example of non-wear time during the day (between 12:00 and

14:00) and we can see how the HSMM approach compares to the traditional cut points approach in application to a sample of accelerometer data.

Figure 2 (C) shows the time spent in each of the HSMM states by the population of children. We can see that before 06:00 and after 23:00 most children are not wearing the device, but between these times the proportion wearing the device quickly rises to over 80%. There is a dip in the total children wearing the device around mid-day, likely to be attributable to nap time. The most active states (4 and 5) comprise about 20% of time spent throughout the day.

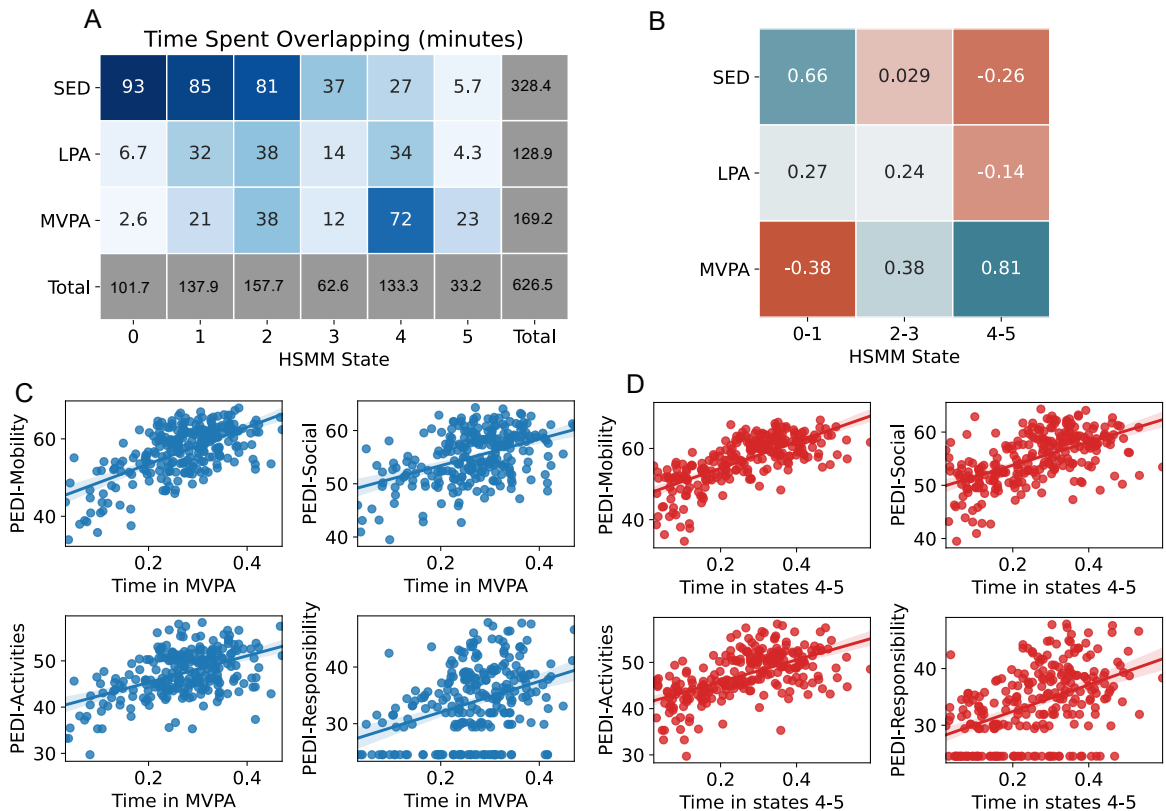

*Figure 3: (A) Shows the mean time (in minutes) that each cut point derived category spends overlapping with each HSMM state. The states are sorted so that state 0 has the lowest mean acceleration and state 5 the highest. (B) Shows the correlation (Spearman's rho) between the time spent in each of the traditional cut points based physical activity intensity categories compared with the time spent in the grouped states of the HSMM. (C) Shows the linear regression of time spent doing MVPA against each of the PEDI-CAT domains. (D) Shows the linear regression of time spent in states 4 and 5 of the HSMM against each of the PEDI-CAT domains.*

Figure 3 (A) shows the extent to which the output of the traditional cut points approach overlaps with that of the HSMM approach. Figure 3 (B) also indicates the extent of agreement between approaches, showing the correlation between outputs. We can see that SED and MVPA are strongly correlated with states 0-1 (rho=0.67), and 4-5 (rho=0.82), respectively. SED is negatively correlated with states 4-5 (rho=-0.28) and MVPA negatively correlated with states 0-1 (rho=-0.39). LPA does not show strong correlation with any state grouping.

To further assess the clinical validity of each approach, we compared the outputs to the four domains of the Paediatric Evaluation of Disability Computer Aided Test (PEDI-CAT) applied to the children from our sample. Figure 3 (C) shows the linear regression of time spent doing MVPA (as a proportion of time wearing the device) and the score achieved by the child in each of the four PEDI-CAT domains. Figure 3 (D) shows the same but for the time spent in the HSMM model states 4 and 5. Table 2 shows the coefficient of determination calculated from the above regression analyses, as well as a regression with age. The output of HSMM explains more variance in each of the four PEDI-CAT domains and age than the time spent in MVPA.

Table 2: Shows the coefficient of determination ($R^2$) calculated on the linear regression of the time spent in HSMM states 4 and 5 or time spent in MVPA with age, and the four domains of the PEDI-CAT measure of ability.

|  | $R^2$ for each method | |
| --- | --- | --- |
| Measure | Cut Points (MVPA) | HSMM (states 4 – 5) |
| Age | 0.09 | 0.14 |
| PEDI Mobility | 0.39 | 0.51 |
| PEDI Social Cognitive | 0.20 | 0.33 |
| PEDI Activities | 0.24 | 0.34 |
| PEDI Responsibility | 0.13 | 0.21 |

## Discussion

In the study reported, we have shown that hidden semi-Markov models can be used to segment and cluster multiday accelerometer data recorded in an ability-diverse population of children aged 9 to 36 months. We have further shown that the resulting activity states are better predictors of the children's developmental capacity than the traditional analytical approach of using cut points.

Previous work (11) has shown that the HSMM can be used to segment and cluster wrist-worn accelerometer data recorded in teenagers. We have built on this by demonstrating that the method can be used in young children (a rapidly developing and physically diverse population) and show that it better captures their movement capacities than the traditional cut points approach.

The traditional cut points method used to quantify physical activity intensity from accelerometry relies on parameters derived from previous studies on external populations (22). If these populations are not representative of those the method is applied to, there is a risk that the parameters may not be calibrated appropriately. This can lead to the method failing to capture enough of the variance in physical activity behaviour present in the sample. For example, if the threshold for MVPA is calculated from a population of physically able adults and then applied to children with physical disabilities, the method would consider most of the population inactive most of the time. While they might rarely reach the physical activity intensities of able adults, there could well be important variations in this population that have not been detected. This can in part be overcome by calculating thresholds on the appropriate populations (7) however, this is expensive and is not feasible in all populations, e.g. in young children or in people with complex disabilities or health conditions. Furthermore, even when appropriate thresholds are available, if one is studying a *diverse* population, or a rapidly evolving population, thresholds for different subpopulations or timepoints would be required, making it difficult to compare between groups and over time. This is a major challenge for longitudinal studies of young children or studies involving participants with a diverse range of physical activity behaviours – often resulting in exclusion of these populations from research.

In contrast to the cut points approach, the HSMM learns the parameters for its hidden states from the data given to it, and the states can then be used to quantify and describe physical activity. This approach has several potential advantages over the cut points approach. The parameters generated by HSMM can be easily interpreted – the mean of the Gaussian distribution representing the observation distribution for the state indicating the physical intensity of the state and the λ parameter of the Poisson distribution representing its duration. The time spent by a participant in each hidden state quantifies their physical activity participation. Our results further suggest the HSMM approach also has better clinical validity compared to the cut points, as the estimates of physical activity participation produced by the HSMM approach correlated more strongly with children's developmental capacity than the cut points estimates. Furthermore, it has the potential to make movement and physical activity research more inclusive for populations where calibration to energy expenditure is not possible – potentially reducing health inequalities over time.

There are two key considerations to using the HSMM approach. First, it requires multiple iterations of its learning algorithm over the entire training dataset, and related high performance computing facilities and programming expertise. Second, the output cannot be directly associated with energy expenditure and thus its best suited for studies where direct estimation of energy expenditure is not a primary concern – such as in developmental studies with young children where movement and physical activity per se are the primary focus, or where additional measures are used to assess subsequent health and biological outcomes.

## Acknowledgements

Dr Niina Kolehmainen, ICA Senior Clinical Lecturer Fellow, NIHR ICA-SCL-2015-01-003, is funded by Health Education England (HEE) / National Institute for Health Research (NIHR) for this research project.

Dr Christopher Thornton, Research Associate, is funded by the National Institute for Health Research (NIHR ICA-SCL-2015-01-003) for this project.

The work of KN is supported by EPSRC, UK under grant EP/R004242/2.

## References

1. Carson V, Lee EY, Hewitt L, Jennings C, Hunter S, Kuzik N, et al. Systematic review of the relationships between physical activity and health indicators in the early years (0-4 years). BMC Public Health. 2017;17.

2. Janssen X, Martin A, Hughes AR, Hill CM, Kotronoulas G, Hesketh KR. Associations of screen time, sedentary time and physical activity with sleep in under 5s: A systematic review and meta-analysis. Sleep Med Rev. 2020 Feb 1;49:101226.

3. Kim Y, Beets MW, Welk GJ. Everything you wanted to know about selecting the "right" Actigraph accelerometer cut-points for youth, but...: A systematic review. Vol. 15, Journal of Science and Medicine in Sport. J Sci Med Sport; 2012. p. 311–21.

4. Lynch BA, Kaufman TK, Rajjo TI, Mohammed K, Kumar S, Murad MH, et al. Accuracy of Accelerometers for Measuring Physical Activity and Levels of Sedentary Behavior in Children: A Systematic Review. Vol. 10, Journal of Primary Care and Community Health. J Prim Care Community Health; 2019.

5. Byrne NM, Hills AP, Hunter GR, Weinsier RL, Schutz Y. Metabolic equivalent: One size does not fit all. J Appl Physiol. 2005;99(3):1112–9.

6. Hildebrand M, Van Hees VT, Hansen BH, Ekelund U. Age group comparability of raw accelerometer output from wrist-and hip-worn monitors. Med Sci Sports Exerc. 2014;46(9):1816–24.

7. Oftedal S, Bell KL, Davies PSW, Ware RS, Boyd RN. Validation of accelerometer cut points in toddlers with and without cerebral palsy. Med Sci Sports Exerc. 2014;46(9):1808–15.

8. Airaksinen M, Räsänen O, Ilén E, Häyrinen T, Kivi A, Marchi V, et al. Automatic Posture and Movement Tracking of Infants with Wearable Movement Sensors. Sci Rep. 2020 Jan 13;10(1):1–13.

9. Albert M V, Sugianto A, Nickele K, Zavos P, Sindu P, Ali M, et al. Hidden Markov model-based activity recognition for toddlers. Physiol Meas. 2020 Mar 6;41(2).

10. Jones P, Mirkes EM, Yates T, Edwardson CL, Catt M, Davies MJ, et al. Towards a portable model to discriminate activity clusters from accelerometer data. Sensors (Switzerland). 2019 Oct 2;19(20).

11. Van Kuppevelt D, Heywood J, Hamer M, Sabia S, Fitzsimons E, Van Hees V. Segmenting accelerometer data from daily life with unsupervised machine learning. PLoS One. 2019 Jan 1;14(1):e0208692.

12. Van Kuppevelt D, Heywood J, Hamer M, Sabia S, Fitzsimons E, Van Hees V. Segmenting accelerometer data from daily life with unsupervised machine learning. PLoS One. 2019;14(1):1–19.

13. Kolehmainen N, Rapley T, Pearce MS. ActiveCHILD Protocol [Internet]. 2017. Available from: https://data.ncl.ac.uk/account/projects/80552/articles/17143841

14. Kolehmainen N, Craw O, Thornton C, Rapley T, Van Sluijs E, Kudlek L, et al. Physical activity in young children across developmental abilities and health states: a cross-sectional analysis of the ActiveCHILD cohort. BMC Med (To be sumbitted Jan 2022). 2022;

15. Nuffield Council on Bioethics (UK). Children and Clinical Research: Ethical Issues [Internet]. Nuffield Council on Bioethics (UK). 2015. Available from: https://www.nuffieldbioethics.org/publications/children-and-clinical-research

16. Kolehmainen N. ActiveCHILD Instructions for wearing the accelerometer [Internet]. 2017. Available from: https://data.ncl.ac.uk/account/projects/80552/articles/17143940

17. Van Cauwenberghe E, Gubbels J, De Bourdeaudhuij I, Cardon G. Feasibility and validity of accelerometer measurements to assess physical activity in toddlers. Int J Behav Nutr Phys Act. 2011;8(1):67.

18. Dumas HM, Fragala-Pinkham MA, Haley SM, Ni P, Coster W, Kramer JM, et al. Computer adaptive test performance in children with and without disabilities: Prospective field study of the PEDI-CAT. Disabil Rehabil. 2012;34(5):393–401.

19. Bakrania K, Yates T, Rowlands A V., Esliger DW, Bunnewell S, Sanders J, et al. Intensity thresholds on raw acceleration data: Euclidean norm minus one (ENMO) and mean amplitude deviation (MAD) approaches. PLoS One. 2016 Oct 1;11(10):e0164045.

20. Penpraze V, Reilly JJ, MacLean CM, Montgomery C, Kelly LA, Paton JY, et al. Monitoring of Physical Activity in Young Children: How Much Is Enough? Pediatr Exerc Sci. 2006 Nov 1;18(4):483–91.

21. Johnson MJ, Willsky AS. Bayesian nonparametric Hidden semi-Markov models. J Mach Learn Res. 2013;14(1):673–701.

22. Pate RR, Almeida MJ, McIver KL, Pfeiffer KA, Dowda M. Validation and calibration of an

accelerometer in preschool children. Obesity. 2006 Nov;14(11).

23. Janssen X, Cliff DP, Reilly JJ, Hinkley T, Jones RA, Batterham M, et al. Predictive validity and classification accuracy of actigraph energy expenditure equations and cut-points in young children. PLoS One. 2013 Nov 11;8(11).

24. Migueles JH, Rowlands A V., Huber F, Sabia S, van Hees VT. GGIR: A Research Community–Driven Open Source R Package for Generating Physical Activity and Sleep Outcomes From Multi-Day Raw Accelerometer Data. J Meas Phys Behav. 2019 Sep 25;2(3):188–96.